\documentclass{isipta2025}
\usepackage{tikz} % for drawing diagrams
\usepackage{booktabs} % for nice tables
\usepackage{fvextra} % for \Verb in tabular environments
\usepackage{array} % for \multicolumn in tabular environments
\usepackage{tabularx} % for a bit more freedom in table column widths
\usepackage[shortcuts]{extdash}

\title{Function-coherent gambles}

\author[1]{Gregory Wheeler}

\affil[1]{Frankfurt School of Finance \& Management\\Germany}

\begin{document}

\maketitle

%%%%%%%%%%%%%%%%%%%%%%%%%%%%%%%%%%%%%%%%%%%%%%%%%%
%             Abstract and keywords              %
%   *only* for papers, not for poster abstracts  %
%%%%%%%%%%%%%%%%%%%%%%%%%%%%%%%%%%%%%%%%%%%%%%%%%%

\begin{abstract}% No length restrictions, but keep things reasonable
	The desirable gambles framework provides a foundational approach to imprecise probability theory but relies heavily on linear utility assumptions. This paper introduces {\em function-coherent gambles}, a generalization that accommodates non-linear utility while preserving essential rationality properties. We establish core axioms for function-coherence and prove a representation theorem that characterizes acceptable gambles through continuous linear functionals. The framework is then applied to analyze various forms of discounting in intertemporal choice, including hyperbolic, quasi-hyperbolic, scale-dependent, and state-dependent discounting. We demonstrate how these alternatives to constant-rate exponential discounting can be integrated within the function-coherent framework. This unified treatment provides theoretical foundations for modeling sophisticated patterns of time preference within the desirability paradigm, bridging a gap between normative theory and observed behavior in intertemporal decision-making under genuine uncertainty.
\end{abstract}


%%%%%%%%%%%%%%%%%%%%%%%%%%%%%%%%%%%%%%%%%%%%%%%%%%
%              Content of the paper              %
%%%%%%%%%%%%%%%%%%%%%%%%%%%%%%%%%%%%%%%%%%%%%%%%%%

\section{Introduction}

The desirable gambles framework \cite{Williams:1975,Walley:2000,DeCoomanQuaeghebeur:2012,Wheeler:2022,DeBock:2023,deCoomanVanCampDeBock:2023} is the  elemental core of a general account of imprecise probabilities, one that includes as a special case the theories of lower  \cite{Walley:1991,LowerPrevisions} and linear previsions \cite{DeFinetti:1974}. Nevertheless, for all its generality, the theory relies on linear utility, a commitment made explicit by the coherence axioms for desirable gambles. Apart from mathematical convenience, which should never be mistaken for a normative principle, the case for linear utility is thin \cite{DeFinetti:1974,Wheeler:2018-sep}.  Indeed, the notion that the utility of satisfying one's desires is non-linear is as old as the concept of utility itself \cite{Bernoulli:1738}. %Behavioral research has long demonstrated systematic deviations from constant-rate exponential discounting in both human and animal inter-temporal choice \cite{frederick2002,Wheeler:2018-sep}.

Recent work \cite{CasanovaBenavoliZaffalon:2021,Wheeler:2021.isipta,Benavoili:2023} has focused on relaxing this linearity commitment within desirable gambles, complementing studies of almost linear uncertainty measures \cite{Corsato:2019}.  Unlike \cite{CasanovaBenavoliZaffalon:2021,Benavoili:2023}, which explore geometric and classification-theoretic formulations of non-linear desirability, the approach in  \cite{Wheeler:2021.isipta} and here maintains an axiomatic foundations and connects non-linear utility to representation via continuous functionals.  In \cite{Wheeler:2021.isipta},  two routes were initially explored. The first preserves the additive structure of the desirable gambles framework and the machinery for coherent inference but detaches the interpretation of desirability from the multiplicative scale invariance axiom. This conservative amendment introduces an adjustable utility scale to discount gambles while maintaining the underlying coherence properties that make the framework effective. The second approach strays from the additive combination axiom altogether to accommodate repeated gambles that return rewards by a non-stationary process. %Although this radical departure introduces non-additive dynamics in sequential gambles, the additive structure appears as a special case. 
Common to both is a method for describing rewards called {\em discounted utility}, which includes linear utility as a special case.

This paper builds on discounted utility \cite{Wheeler:2021.isipta} by introducing {\em function-coherent gambles} that preserve the core rationality requirements of the desirable gambles approach while relaxing the commitment to linear utility. We focus particularly on applications to discounting and intertemporal choice, where linear utility assumptions have proven especially restrictive. We examine several well-documented alternatives to constant-rate exponential discounting: hyperbolic discounting, which exhibits steeper drops for near-term outcomes; quasi-hyperbolic discounting, incorporating immediate preference for present consumption; scale-dependent discounting for magnitude effects; state-dependent discounting for varying environmental conditions; and hybrid approaches combining multiple forms. For each, we analyze their mathematical properties, empirical support, and implications for coherence and inference.  In a separate paper \cite{Wheeler:2025.dynamics}, we relax additive combination to accommodate repeated gambles played over time where rewards compound multiplicatively.

The paper proceeds as follows. Section 2 reviews the standard coherence framework and introduces basic notation. Section 3 presents the axioms for function-coherent gambles and establishes their key properties. Section 4 proves a representation theorem  that characterizes acceptable gambles through continuous linear functionals. Section 5 explores connections to risk measures, while Section 6 applies the framework to various forms of discounting.

\section{Preliminaries}

The desirable gambles framework consists of axioms for constructing a coherent set $\mathbb{D} \subseteq X$ of bounded gambles directly from an initial collection \cite{Williams:1975,Walley:2000,IntroIP}. For bounded gambles $f, g$ and positive real number $\lambda \geq 0$, a set of bounded gambles $\mathbb{D}$ is coherent when satisfying:

\begin{itemize}
\item[A1.]  If $f < 0$, then $f \not\in \mathbb{D}$ \hfill ({\bf Avoid partial loss})
\item[A2.] If $f \geq 0$, then $f \in \mathbb{D}$  \hfill ({\bf Accept partial gain}) 
\item[A3.] If $f \in \mathbb{D}$, then $\lambda f \in \mathbb{D}$ \hfill ({\bf Pos.~Scale Invariance})
\item[A4.]  If $f \in \mathbb{D}$ and $g \in \mathbb{D}$, then $f + g \in \mathbb{D}$  \hfill ({\bf Combination})
\end{itemize}
\noindent Gambles are assessed pointwise. For instance, partial loss avoidance states $f \not\in \mathbb{D}$ if $f(s) < 0$, for every state $s$.

Axioms A1 and A2 express rationality conditions, while A3 and A4 are closure operations encoding linear utility. Together they define a convex cone containing all conic combinations of its elements:

\begin{equation}
\label{eq:conic_hull}
\mathsf{cone}(\mathbb{D}) := \left\{ \sum_{i =1}^{n} \lambda_i f_i :  f_i\in \mathbb{D}, i=1,\ldots,n,\; \lambda_i \geq 0   \right\}
\end{equation}

For a gamble $f$ over $|\Omega|=m$ states, the reward vector $\mathbf{x}_f = (x_1, \ldots, x_m)$ in $\mathbb{R}^m$ represents state-contingent outcomes, with linear utility:

\begin{equation}
\label{eq:lin-U}
u_{lin}(\mathbf{x}_f) = \mathbf{1}\mathbf{x} = \mathbf{x}
\end{equation}

The baseline discounted utility model \cite{Wheeler:2021.isipta} introduced to accommodate non-linear utility is:

\begin{definition}[Discounted Utility]
\label{df:utility_discounted_original}
%Let $\alpha \in [0, 1)$ and $x >0$. 
\begin{equation}
\label{eq:util_original}
u(x,\alpha) := \frac{x^{1-\alpha}-\alpha}{1-\alpha}, \quad \alpha \in [0,1),\; x > 0.
\end{equation}
\end{definition}

\noindent Discounted utility takes a positive scalar reward $x$ and discounts the desirability of $x$ to degree $\alpha$.  When no discounting is applied, $\alpha = 0$, then the utility of $x$ is linear, $u(x,\alpha) = x$.  Alternatively, discounted utility approximates the natural logarithm of $x$ as $\alpha$ approaches $1$.

Discounted utility requires positive rewards, accommodated by assuming gambles modify a current bank of positive wealth, $w$, sufficient to cover potential losses. Unlike linear utility, where stake size is irrelevant, size matters fundamentally to discounted utility.  Thus, we assume agents are only presented with gambles bounded below by $-w$.

\section{Function-Coherent Gambles}

The standard coherence axioms encode linear utility through scale invariance (A3) and additive combination (A4). However, for non‐linear utility, these operations do not preserve the reward structure. We therefore introduce axioms that maintain a weaker notion of coherence under utility transformations that satisfy basic rationality requirements.  The discounted utility in Definition~\ref{df:utility_discounted_original}  satisfies all the requirements for admissible utility functions here and provides a motivating special case of function-coherence.

Let $\mathbb{D}$ be a set of acceptable gambles\footnote{Here I use `desirable'  to refer to coherent desirable gambles  with respect to axioms A1--A4 and introduce `acceptable' to discuss function-coherent gambles with respect to axioms F1--F3.  Outside the differences between these two axiomatic systems and their consequences,  acceptability and desirability are treated as operationally equivalent.} and $u: X \to \mathbb{R}$ a strictly increasing, continuous utility function on domain $X$ with normalization $u(0) = 0$. A set $\mathbb{D} \subseteq X$ is \emph{function-coherent} when satisfying:
 
\begin{enumerate}[label=(F\arabic*)]
    \item \textbf{Avoid Partial Losses:} If $f < 0$, then $f \notin \mathbb{D}$.
    \item \textbf{Monotonicity:} If $f \ge g$ and $g \in \mathbb{D}$, then $f \in \mathbb{D}$.
    \item \textbf{$u$-Convexity:} For $f, g \in \mathbb{D}$ and nonnegative $\lambda, \mu$ where 
    $$h = u^{-1}\Bigl(\lambda\, u(f) + \mu\, u(g)\Bigr)$$ 
    is defined, $h \in \mathbb{D}$.
\end{enumerate}

As for the well-definedness of $u^{-1}$ in F3, we make two assumptions for admissible utility functions:

\begin{enumerate}[align=left, leftmargin=*, labelwidth=!, labelsep=0.5em]
\item[(F3a)] The utility function $u: X \to V$ is a strictly increasing and continuous bijection onto its image, $u(X) \subseteq V$.

\item[(F3b)] The image $u(X)$ is convex.  That is, for any $f, g \in \mathbb{D}$ and any nonnegative scalars $\lambda, \mu$, the linear combination $\lambda u(f) + \mu u(g)$ is always within the domain of $u^{-1}$.
\end{enumerate}

Standard coherence emerges when $u$ is the identity function. In what follows, acceptance is defined by $u(f) \ge 0$ and $u(0)=0$.  We now turn to stating four key properties.  

\begin{proposition}[Non-Triviality]
Under F1 and F2:
\begin{enumerate}
    \item $\mathbb{D}$ is nonempty and contains no sure losses,
    \item Every $f$ with $u(f) \ge 0$ is acceptable.
\end{enumerate}
\end{proposition}

\begin{proposition}[Upward Closure]
If $f \in \mathbb{D}$ and $g \in X$ satisfies $g(s) \ge f(s)$ for all states $s$, then $g \in \mathbb{D}$.
\end{proposition}

\begin{proposition}[Transform Convexity]
The $u$\=/trans\-/formed set $U(\mathbb{D}) := \{ u(f) : f \in \mathbb{D} \}$ is a convex cone under F3.
\end{proposition}

\begin{proposition}[Transform Invariance]
For strictly increasing $\phi$ with $\phi(0)=0$, if $\tilde{u} = \phi \circ u$ then
$$\{ f \in X : \tilde{u}(f) \ge 0 \} = \{ f \in X : u(f) \ge 0 \} = \mathbb{D}.$$
\end{proposition}

\section{Representation}

Let $X$ be a real vector space of gambles, $V$ a locally convex, Hausdorff topological vector space of utilities or valuations, and $u: X \to V$ a strictly increasing, continuous function that transforms a gamble into a scalar representation of preference.  Consider the acceptance set
$$\mathbb{D} = \{ f \in X : u(f) \ge 0 \},$$
normalized so $u(0)=0$. Under function-coherence (F1--F3) and the following regularity conditions:
\begin{enumerate}[label=(R\arabic*)]
    \item $U(\mathbb{D})$ is closed,
    \item $U(\mathbb{D})$ possesses a non-empty interior in $V$,
\end{enumerate}
separation theorems can be applied in the proof of the representation theorem below.

As assumed above,  $u$ is a strictly increasing, continuous bijection onto its image.  If  $u(X)$ is convex and any convex combination $\lambda u(f) + \mu u(g)$ lies in the domain of $u^{-1}$, then the following representation holds:

\begin{theorem}[Representation]
\label{thm:rep}
There exists a continuous linear functional $\ell: V \to \mathbb{R}$, unique up to positive scaling, such that
$$f \in \mathbb{D} \quad \Longleftrightarrow \quad \ell(u(f)) \ge 0.$$
\end{theorem}
\noindent Here $u$ encodes preferences and $\ell$ captures belief aggregation, a point discussed further in Section~\ref{risk-measure}.

Next, suppose $X$ is a topological vector space and the utility function $u: X \to \mathbb{R}$ is continuous.  Then, $\mathbb{D}$ is closed in this topological space. 
\begin{theorem}[Closure Under Limits]
An acceptance set is $\mathbb{D} = \{f \in X : u(f) \geq 0 \}$. If $\{f_n\}$ is a sequence in $\mathbb{D}$ that converges to some $f \in X$, then $f \in \mathbb{D}$.
\end{theorem}

\section{Representation and Risk Measures}
\label{risk-measure}

The representation theorem \ref{thm:rep} connects together function-coherent acceptance sets and continuous linear functionals through the risk functional 
$$\rho(f) = \ell(u(f)).$$
\noindent This connection has three important implications:

\subsection*{Risk Assessment Structure}

The functional $\rho$ is a generalized risk measure incorporating non-linear utility. A generalized risk measure is a real-valued functional, $\rho$, on a set of gambles satisfying monotonicity and convexity. (See \cite{FollmerSchied:2002} for a foundational treatment.)  Unlike traditional risk measures that operate directly on gambles, $\rho$ first transforms outcomes through $u$, capturing an agent's subjective valuation, before aggregating with $\ell$. For example, when $u$ is concave ($\alpha > 0$, which reflects the degree of risk aversion in the utility curvature), $\rho$ naturally encodes risk aversion.

Consider two functionals $\rho_1$ and $\rho_2$ defined by different concave utilities $u_1$ and $u_2$. Even with identical $\ell$, these functionals induce different orderings over gambles, reflecting how risk perception depends fundamentally on utility structure. This stands in contrast to linear risk measures where utility assumptions are typically implicit.

\subsection*{Preference-Belief Decomposition} 

The composition $\ell \circ u$ cleanly separates:
\begin{itemize}
    \item {\em Preferences} encoded by utility $u$, and 
    \item {\em Beliefs} captured by aggregator $\ell$.
\end{itemize}

This decomposition enables one to analyze how changes in risk attitudes (via $u$) interact with changes in belief weighting (via $\ell$). For instance, increasing risk aversion by adjusting $u$ affects valuation independently of how states are aggregated by $\ell$.

\subsection*{Ordering Invariance}

The uniqueness of $\ell$ up to positive scaling has an important implication. If $\ell_1$ and $\ell_2$ both satisfy
\[f \in \mathbb{D} \iff \ell_i(u(f)) \ge 0,\]
then there exists $c > 0$ such that
\[\ell_2 = c\ell_1.\]

While absolute risk values may differ between representations, the induced ordering remains invariant. Thus,  relative preferences, not absolute measures, determine choice behavior.

\begin{example}
Return to $X = \mathbb{R}^2$ with $u(x) = \log(1 + x)$, gambles $f_1, f_2$, and weights $(w_1, w_2)$. Assume $f_i > -1$ to ensure utility is well-defined. The risk functional
$$\rho(f) = w_1\log(1 + f_1) + w_2\log(1 + f_2)$$
exhibits all three properties:
\begin{enumerate}
    \item Risk aversion through concave $u$,
    \item Separation of utility ($\log$) from beliefs $(w_1, w_2)$,
    \item Invariance to rescaling weights.
\end{enumerate}
\end{example}

\section{Function-Coherent Discounting}

The analysis of intertemporal choice depends on how we model the relationship between time delay and reward valuation. One approach is to  construct an {\em effective utility function} of the form:
$$
v(x,t)=u\bigl(D(t)x\bigr)
$$
where $u$ represents the agent's utility function, $x$ denotes the reward magnitude, and $D(t)$ captures the discount factor for delay $t$. This composition structure provides a flexible framework for modeling time preference while maintaining essential normative attributes of coherent desirability.

A key  advantage of this use of effective utility  lies in its preservation of monotonicity properties. For any fixed delay $t$, the discount factor $D(t)$ acts as a positive scaling constant, ensuring that the transformation $x\mapsto D(t)x$ remains strictly increasing and continuous. When composed with a strictly increasing and continuous utility function $u$, the resulting effective utility $v(x,t)$ inherits these properties. Consequently, the acceptance set
$$
\mathbb{D}_t=\{f\in X: v(f,t)\ge0\}
$$
satisfies the fundamental axioms of function-coherence (F1)--(F3), providing  foundations for more sophisticated models of time preference.

The function-coherent framework accommodates various specifications of the discount function $D(t)$, each capturing distinct aspects of intertemporal choice behavior. Standard exponential discounting, where $D(t)=e^{-rt}$, provides a benchmark case that ensures time consistency but often fails to capture empirically observed patterns of choice \cite{Wheeler:2018-sep}. Alternative specifications can incorporate phenomena such as diminishing impatience, present bias, magnitude effects, and state dependence while maintaining the essential dominance structure of coherent choice.

In this section we examine a range of discount functions through the lens of function-coherence.  To illustrate these effects, we consider positive rewards. Each specification illustrates how sophisticated patterns of time preference are handled within the desirability framework by maintaining the composition structure $u(D(t)x)$ while varying the form of $D(t)$.

We proceed from simpler to more complex discounting functions,  beginning with hyperbolic discounting. Hyperbolic discounting  introduces diminishing impatience through a single parameter. We then examine quasi-hyperbolic discounting, which separates present bias from long-run patience into two local components. Subsequent sections explore generalizations that incorporate magnitude effects, state dependence, and hybrid structures. Throughout, we will  focus on how each specification preserves function-coherence while capturing distinct aspects of intertemporal choice behavior.

This systematic examination serves multiple purposes. First, it demonstrates the flexibility of the function-coherent framework in accommodating diverse patterns of time preference. Second, it provides theoretical foundations for using alternative discount functions within a desirability setting. Finally, it illustrates how apparently distinct approaches to modeling time preference can be unified through the common structure of function-coherence.

\subsection{Hyperbolic Discounting}

Hyperbolic discounting is a theoretically grounded approach to modeling observed patterns of time preference, particularly the tendency for individuals to exhibit steeper discounting over near-term horizons while displaying greater patience for long-term outcomes \cite{LoewensteinPrelec:1992,Mazur:1987,Harris:2013}. This discount function takes the form:
$$
D_H(t)=\frac{1}{1+kt},\quad k>0,
$$
where $k$ controls the rate of time preference. For any fixed delay $t\ge0$, $D_H(t)$ remains strictly positive, ensuring that the effective utility
$$
v(x,t)=u\Bigl(\frac{x}{1+kt}\Bigr)
$$
preserves monotonicity in $x$ and maintains function-coherence within each temporal frame.

The hyperbolic structure generates systematically different patterns of discounting compared to the standard exponential framework. The discount rate at time $t$, given by $k/(1+kt)$, decreases as the horizon extends, capturing diminishing impatience while maintaining analytical tractability. This feature proves particularly valuable for modeling phenomena like preference reversals and dynamic inconsistency within a rational choice framework \cite{Thaler:1981,RegenwetterDanaDavis-Stober:2011}.

\begin{example}[Investment Choice with Time-Driven Preference Reversal]
%Consider an investment manager evaluating two pairs of opportunities that differ solely in their timing. In both cases the later payoff is 20\% larger than the earlier one:
Consider an investment manager evaluating two pairs of opportunities:
\begin{align*}
\text{Project A:} & \begin{cases}
\text{\$1000 immediately} \\
\text{\$1200 in 1 year}
\end{cases} \\[2mm]
\text{Project B:} & \begin{cases}
\text{\$1000 in 5 years} \\
\text{\$1200 in 6 years}
\end{cases}
\end{align*}

Assume a hyperbolic discount function with parameter \(k=0.5\):
\[
D_H(t)=\frac{1}{1+0.5\,t},
\]
and an increasing utility function \(u(x)=\log(1+x)\). The effective utility of a payoff \(x\) received at time \(t\) is given by
\[
v(x,t)= u\Bigl(\frac{x}{1+0.5\,t}\Bigr).
\]

\noindent \textbf{Project A: Near-Term Payoffs}

For the immediate payoff at \(t=0\):
\begin{align*}
v_A(1000,0)&= \log\Bigl(1+\frac{1000}{1+0.5\cdot 0}\Bigr)\\
		&= \log(1001) \approx 6.91.
\end{align*}
For the later payoff at \(t=1\):
\begin{align*}
v_A(1200,1) &= \log\Bigl(1+\frac{1200}{1+0.5\cdot 1}\Bigr)\\
		&= \log\Bigl(1+\frac{1200}{1.5}\Bigr)\\
		&= \log(801) \approx 6.69.
\end{align*}
Thus, when the payoffs are in the near term, the manager prefers the \$1000 received immediately over the \$1200 received in one year.

\noindent \textbf{Project B: Distant Payoffs}

For the payoff at \(t=5\):
\begin{align*}
v_B(1000,5) &= \log\Bigl(1+\frac{1000}{1+0.5\cdot5}\Bigr)\\
&= \log\Bigl(1+\frac{1000}{3.5}\Bigr)\\
&= \log(1+285.7) = \log(286.7) \approx 5.66.
\end{align*}
For the payoff at \(t=6\):
\begin{align*}
v_B(1200,6) &= \log\Bigl(1+\frac{1200}{1+0.5\cdot6}\Bigr)\\
		&= \log\Bigl(1+\frac{1200}{4}\Bigr)\\
		&= \log(1+300) = \log(301) \approx 5.71.
\end{align*}
Now, with both payoffs deferred into the future, the manager prefers the \$1200 in 6 years over the \$1000 in 5 years.

Even though the later payoff is 20\% larger in both projects, hyperbolic discounting—with its declining discount rate over time—causes the agent to prefer the immediate reward in Project A but to reverse this preference in Project B. Time is the driving factor: the only difference between the two projects is the time at which the payoffs are received.
\end{example}

This example illustrates how hyperbolic discounting naturally generates horizon-dependent patience while maintaining function-coherence within each decision context. Rather than relying on behavioral bias, the framework explains preference reversals through systematic variation in discount rates across temporal frames.

For sequences of dated payments $(x_1,t_1),\ldots,(x_n,t_n)$, the total effective utility becomes:
\[
V = \sum_{i=1}^n u\Bigl(\frac{x_i}{1+kt_i}\Bigr)
\]
This formulation maintains analytical tractability while capturing sophisticated patterns of time preference, including diminishing impatience and horizon effects, all within the principles of coherent desirability.

The representation theorem yields a continuous linear functional $\ell$ that characterizes acceptance through:
\[
\ell\Bigl(u\bigl((1+kt)^{-1}x\bigr)\Bigr) \ge 0
\]

\subsection{Quasi-hyperbolic Discounting}

Quasi-hyperbolic discounting \cite{Laibson:1997,ODonoghueRabin:1999}, also known as $\beta$-$\delta$ discounting, provides a tractable approach to modeling present bias while maintaining analytical convenience for future payoffs. The discount function takes the form:
$$
D_Q(t)=
\begin{cases}
1, & \text{if } t=0,\\[1mm]
\beta\,\delta^t, & \text{if } t>0,
\end{cases}
$$
where $\beta\in (0,1]$ captures immediate gratification bias and $\delta\in (0,1)$ governs long-run discounting. This formulation introduces a discrete present bias through an immediate drop in value, followed by standard exponential decay. For any fixed delay $t$, $D_Q(t)$ remains positive, ensuring that the effective utility 
$$
v(x,t)=u\bigl(D_Q(t)x\bigr)
$$
maintains strict monotonicity in $x$.

\begin{figure}%[htb!]

\includegraphics[width=1\columnwidth]{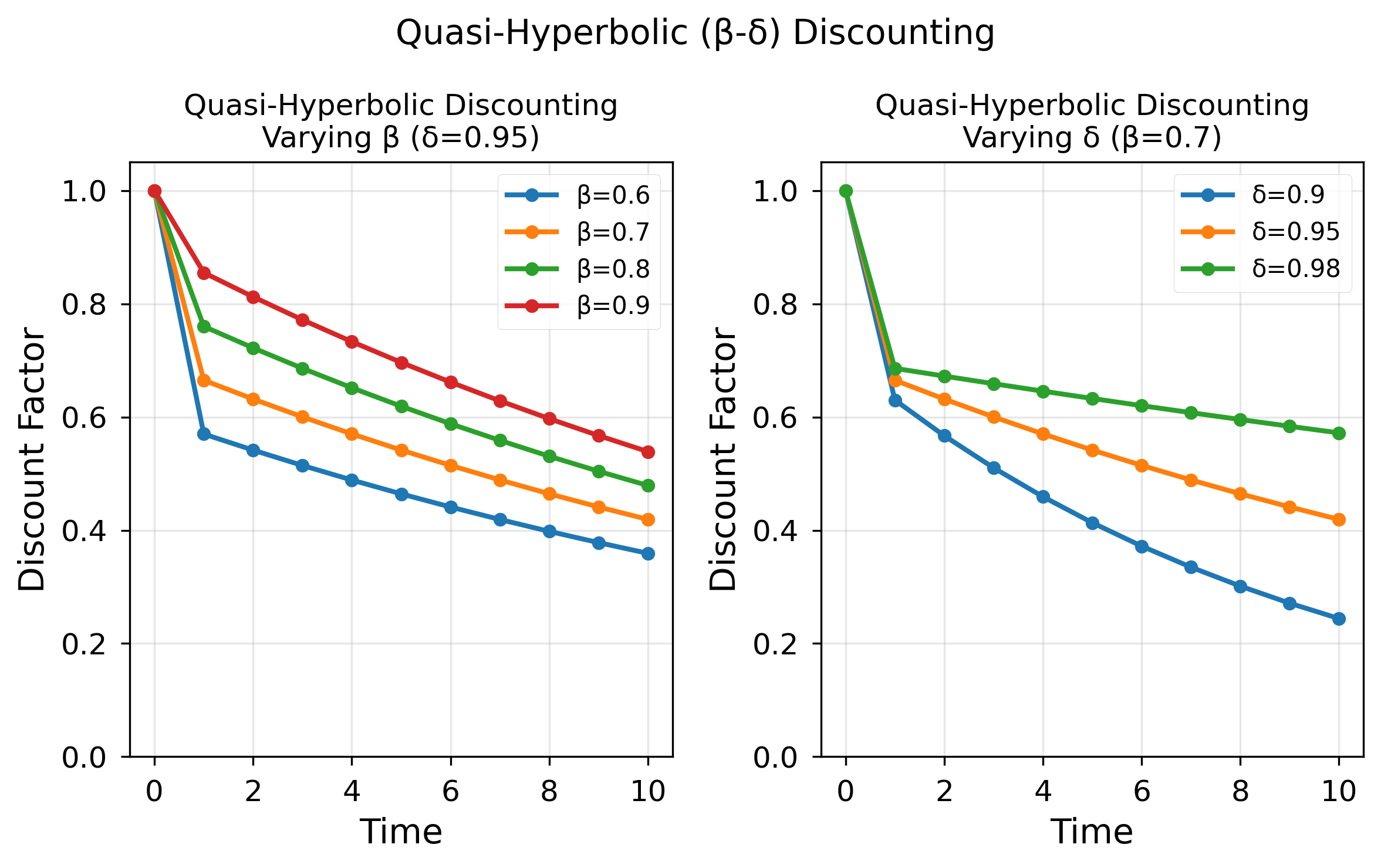}

\caption{{\small Quasi-hyperbolic ($\beta$-$\delta$) discounting with discrete time periods. The left panel shows the effect of varying the present bias parameter $\beta$ (0.6-0.9) while holding the long-run discount factor $\delta$ constant at 0.95. The right panel demonstrates the impact of varying $\delta$ (0.90-0.98) while maintaining $\beta=0.7$. The discontinuity at $t=0$ captures immediate gratification, while subsequent periods follow exponential discounting.}}
\label{fig:quasi_hyperbolic}
\vspace*{-3.0mm}
\end{figure}

The quasi-hyperbolic framework offers distinct advantages in modeling self-control problems and dynamic inconsistency. By separating present bias from long-run patience, it captures sophisticated patterns of time preference while preserving the analytical tractability of exponential discounting for future periods (Fig.~\ref{fig:quasi_hyperbolic}). This structure proves particularly valuable for analyzing phenomena like procrastination and gaps between planned and actual behavior.

\begin{example}[Commitment Savings]
Consider a household choosing between two savings plans:
\begin{align*}
\text{Plan A:} & \begin{cases}
\text{\$100 accessible immediately} \\
\text{\$120 in a 1-month locked account}
\end{cases} \\[2mm]
\text{Plan B:} & \begin{cases}
\text{\$100 accessible in 12 months} \\
\text{\$120 in a 13-month locked account}
\end{cases}
\end{align*}

With present bias parameter $\beta=0.7$, monthly discount factor $\delta=0.95$, and utility $u(x)=\sqrt{x}$, the effective utilities for Plan A become:
\begin{align*}
v_A(100,0) &= \sqrt{100} = 10 \\
v_A(120,1) &= \sqrt{120 \cdot 0.7 \cdot 0.95} \approx 8.93
\end{align*}
indicating a preference for immediate access despite the foregone return.

For the more distant Plan B, the effective utilities are:
\begin{align*}
v_B(100,12) &= \sqrt{100 \cdot 0.7 \cdot 0.95^{12}} \approx 6.15 \\
v_B(120,13) &= \sqrt{120 \cdot 0.7 \cdot 0.95^{13}} \approx 6.57
\end{align*}
revealing a preference reversal that favors committing to the higher return when all payoffs are sufficiently delayed.
\end{example}

%+++++++++
%++++++++

This intertemporal preference reversal emerges naturally from the quasi-hyperbolic structure. When evaluating immediate versus near-term payoffs, present bias dominates decision-making, leading to apparent impatience. However, when comparing two future payoffs, the present bias factor $\beta$ affects both options equally, allowing exponential discounting to drive choice. This mechanism explains empirical patterns like gym membership purchases followed by underutilization: the future self faces different effective utilities than those anticipated by the current self, despite an underlying function-coherence within each temporal frame.

This regime  extends naturally to sequences of decisions, too. Consider a series of savings choices $(s_1,t_1),\ldots,(s_n,t_n)$ with effective utility:
$$
V = \sum_{i=1}^n u\bigl(\beta\mathbf{1}_{t_i>0}\delta^{t_i}s_i\bigr)
$$
where $\mathbf{1}_{t_i>0}$ indicates future payments. The discrete present bias creates a systematic wedge between planned and realized behavior while maintaining analytical tractability through the exponential structure of future discounting.

The representation theorem takes a piecewise form, yielding separate continuous linear functionals $\ell_0$ (present) and $\ell_1$ (future) that characterize acceptance through:
$$
\ell_j\bigl(u(D_Q(t)x)\bigr) \ge 0, \quad j = \mathbf{1}_{t>0}
$$
This dual representation formally captures how present bias creates two distinct but internally coherent decision frameworks, providing theoretical foundations for analyzing behavior that appears inconsistent in aggregate but is nevertheless rational within each temporal segment.

\subsection{Generalized Hyperbolic Discounting}

Generalized hyperbolic discounting extends the standard hyperbolic framework by introducing an additional parameter that controls the evolution of time preferences:
$$
D_G(t)=\frac{1}{(1+kt)^p},\quad k>0,\; p>0
$$
where $k$ establishes the base discount rate and a new parameter, $p$, determines the speed at which preferences change over time. This formulation nests several important special cases: $p=1$ recovers standard hyperbolic discounting, while $p\to\infty$ approximates exponential decay. For any fixed time $t$, $D_G(t)$ remains positive, ensuring that the effective utility
$$
v(x,t)=u\Bigl(\frac{x}{(1+kt)^p}\Bigr)
$$
maintains strict monotonicity in $x$.

\begin{figure}[htb!]

\includegraphics[width=1\columnwidth]{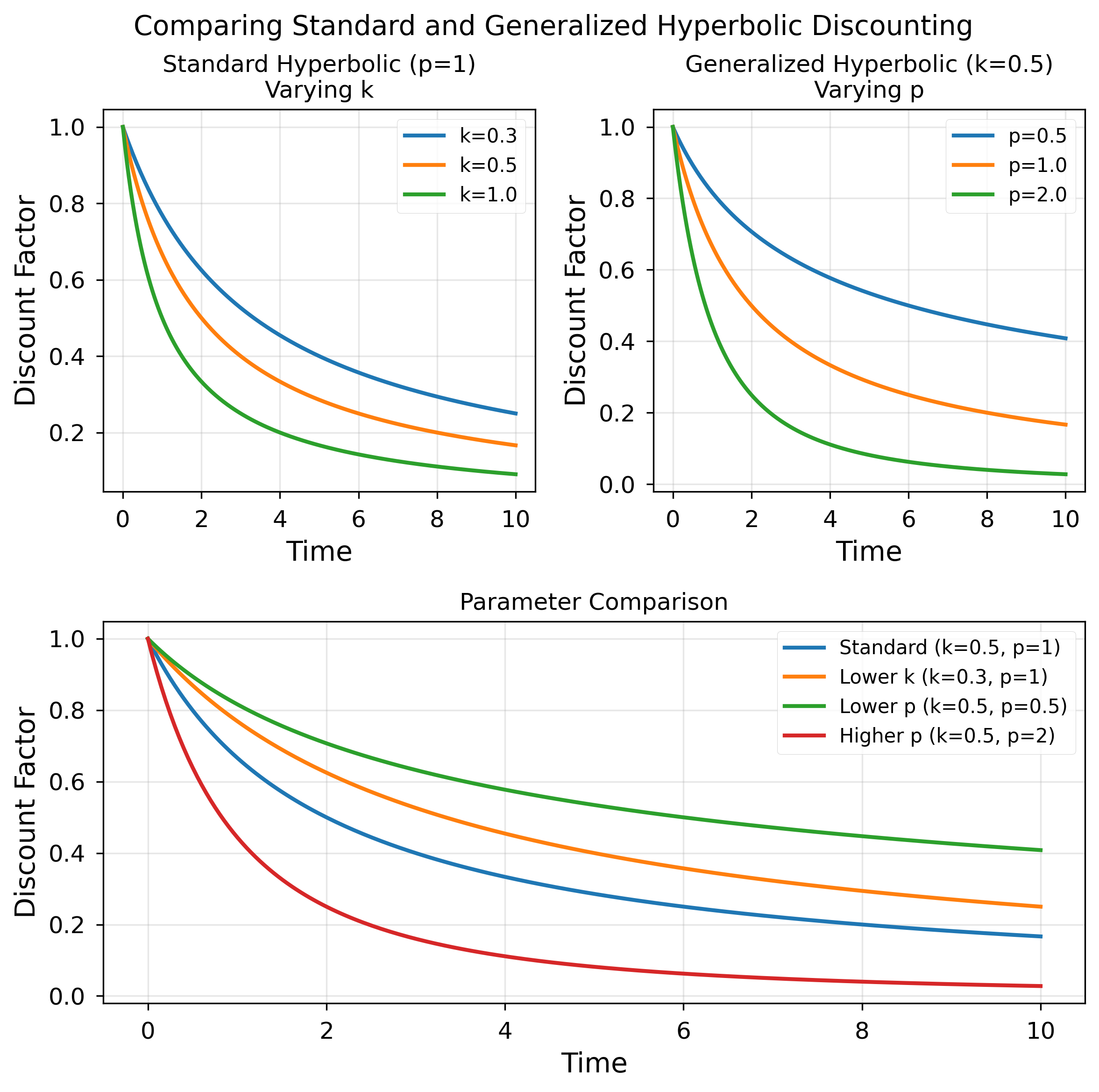}

\includegraphics[width=1\columnwidth]{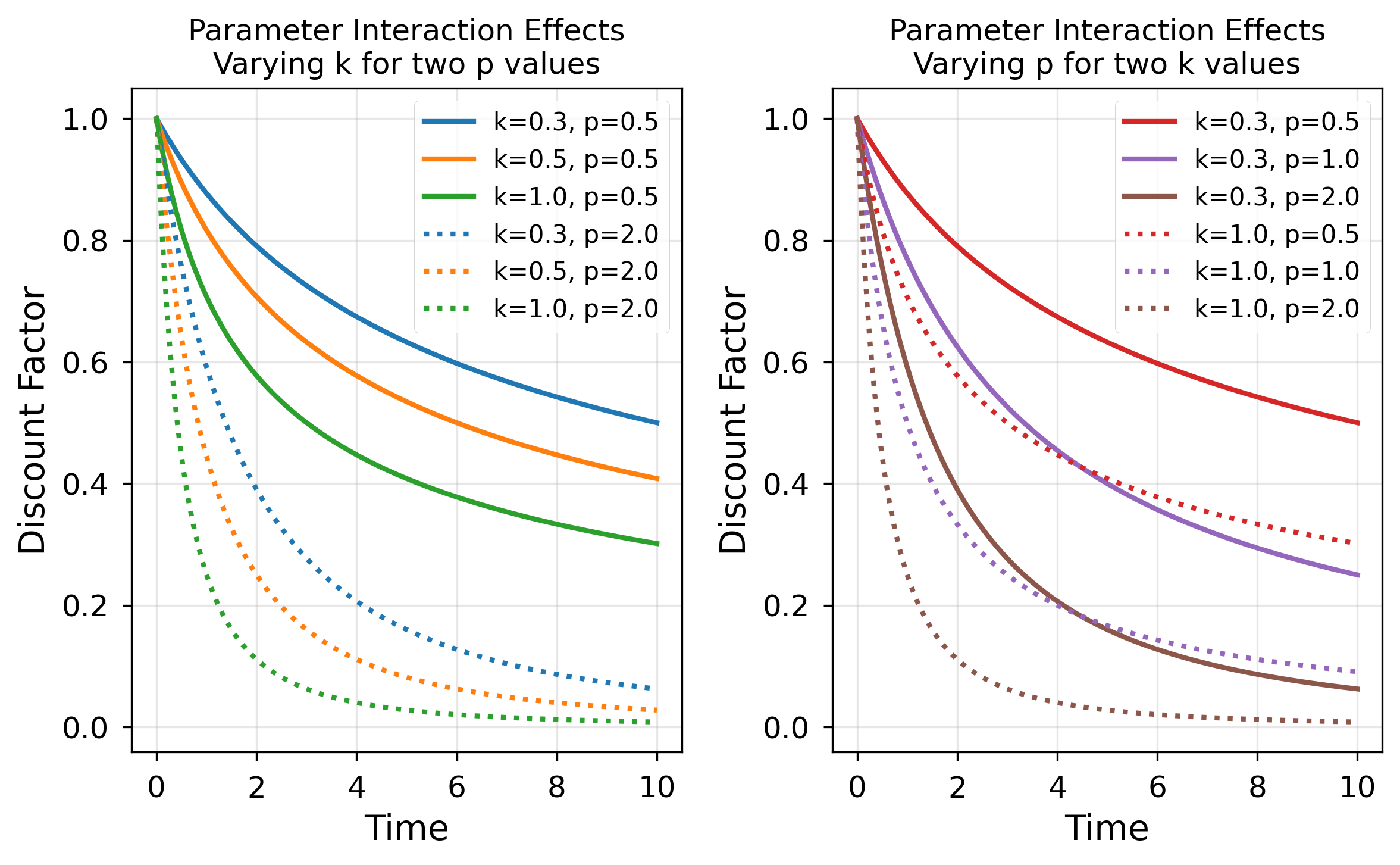}

\caption{{\small Visualization of standard and generalized hyperbolic discounting functions. The top panel shows standard hyperbolic discounting with varying k values (left) and generalized hyperbolic discounting with varying p values (right). The middle row displays a direct comparison of different parameter combinations. The bottom panel demonstrates parameter interaction effects by  varying k values at two different p values (left) and varying p values at two different k values (right). Higher k values produce steeper initial declines, while higher p values affect the tail behavior of the discount function.}}
\label{fig:hyperbolic}
\vspace*{-3.0mm}
\end{figure}

The additional flexibility offered by the power parameter $p$ enables more nuanced modeling of intertemporal preferences. Low values of $p$ generate relatively flat discount curves that exhibit sustained patience over moderate time horizons, while high values of $p$ produce steep initial declines followed by more stable long-run discounting (Fig.~\ref{fig:hyperbolic}). This parametric flexibility proves particularly valuable when modeling heterogeneous time preferences across different decision contexts or populations.

\begin{example}[Investment Choice]
Consider a firm evaluating two investment opportunities:
\begin{align*}
\text{Option A:} & \begin{cases}
\text{\$100K today} \\
\text{\$120K in year 5}
\end{cases} \\[2mm]
\text{Option B:} & \begin{cases}
\text{\$110K in year 2} \\
\text{\$150K in year 4}
\end{cases}
\end{align*}

With base rate $k=0.2$ and linear utility $u(x)=x$, different values of $p$ lead to qualitatively different investment choices. Here we adopt linear utility  to isolate the effects of the discount function. Under relatively patient preferences ($p=0.5$), the discount factors evolve gradually:
\begin{align*}
D_G(2) &= \frac{1}{(1+0.4)^{0.5}} \approx 0.82 \\
D_G(4) &= \frac{1}{(1+0.8)^{0.5}} \approx 0.74 \\
D_G(5) &= \frac{1}{(1+1.0)^{0.5}} \approx 0.71
\end{align*}

These discount factors yield option values:
\begin{align*}
V_A &= 100 + 120(0.71) = 185.2\text{K} \\
V_B &= 110(0.82) + 150(0.74) = 201.2\text{K}
\end{align*}
indicating a preference for Option B's balanced payment structure.

However, more impatient preferences ($p=2$) generate rapidly declining discount factors:
\begin{align*}
D_G(2) &= \frac{1}{(1+0.4)^2} \approx 0.51 \\
D_G(4) &= \frac{1}{(1+0.8)^2} \approx 0.31 \\
D_G(5) &= \frac{1}{(1+1.0)^2} \approx 0.25
\end{align*}

Under this more pronounced discounting, the option values become:
\begin{align*}
V_A &= 100 + 120(0.25) = 130\text{K} \\
V_B &= 110(0.51) + 150(0.31) = 102.6\text{K}
\end{align*}
now favoring Option A due to its larger immediate payment.
\end{example}

The example illustrates how generalized hyperbolic discounting can capture fundamentally different approaches to intertemporal tradeoffs through variation in the power parameter alone. This parametric flexibility proves particularly valuable when modeling heterogeneous time preferences across different decision contexts, market environments, or demographic groups. Moreover, the framework maintains analytical tractability while accommodating rich patterns of preference evolution.

The representation theorem applies directly to generalized hyperbolic discounting, yielding a continuous linear functional $\ell$ that characterizes acceptance through:
$$
\ell\Bigl(u\bigl((1+kt)^{-p}x\bigr)\Bigr) \ge 0
$$

%+++++++++
%++++++++

\noindent This representation demonstrates how the framework preserves function-coherence while capturing sophisticated patterns of temporal preference. %The ability to modulate preference evolution through the power parameter $p$ provides a theoretically grounded approach to modeling heterogeneous time preferences within the rational choice paradigm.

\subsection{Scale-dependent Discounting}
\label{sec:scale-dependent-discounting}

Scale-dependent discounting \cite{Thaler:1981,Benhabib:2010} captures an important empirical regularity in intertemporal choice: individuals often exhibit greater patience when evaluating larger rewards. This phenomenon, known as the magnitude effect \cite{Thaler:1981}, is formalized through a discount function that varies with the reward size:
$$
D_S(t,x)=D(t)^{\,\eta(x)}
$$
where $\eta:\mathbb{R}^+\to\mathbb{R}^+$ modulates the base discount rate according to the reward magnitude. When using a standard exponential base rate $D(t)=e^{-rt}$, smaller values of $\eta(x)$ reduce the effective discount rate, thereby capturing increased patience for larger rewards.

The introduction of reward-dependent discounting presents unique theoretical challenges within the function-coherent framework. The key technical requirement lies in preserving monotonicity: for the mapping $x \mapsto D(t)^{\eta(x)}x$ to increase strictly, we require:
$$
1 + \eta'(x)x\log(D(t)) > 0
$$
This constraint ensures that the framework maintains function-coherence while capturing empirically observed magnitude effects. The precise specification of $\eta(x)$ thus becomes crucial for both theoretical consistency and empirical applicability (Fig~\ref{fig:general-discounting}).

\begin{example}[Magnitude Effect]

Consider two pairs of opportunities:
\begin{align*}
\text{Choice A:} & \begin{cases}
\text{\$10 immediately} \\
\text{\$15 in 1 month}
\end{cases} \\[2mm]
\text{Choice B:} & \begin{cases}
\text{\$1000 immediately} \\
\text{\$1500 in 1 month}
\end{cases}
\end{align*}

Using scale-dependent discounting $D_{S}(t,x) = e^{(-t/log(x))}$ and linear utility $u(x) = x$, the effective utility becomes:
\[ v(x,t) = u(D_{S}(t,x)x) \]
For the small-scale choice, 
\begin{align*}
v(10,0) &= 10 \\
v(15,1) &= 15 \cdot e^{(-1/\log{(15)})} \approx 8.62
\end{align*}
which yields a higher utility  for A.  For the large-scale choice, 
\begin{align*}
v(1000,0) &= 1000 \\
v(1500,1) &= 1500 \cdot e^{(-1/\log{(1500)})} \approx 1222
\end{align*}
revealing relatively greater patience at the higher magnitude, indicating a preference for B.

\end{example}

Scale-dependent discounting introduces magnitude sensitivity while preserving the basic shape of time preference within each reward level. Unlike hyperbolic or quasi-hyperbolic approaches that modify temporal structure, this framework allows discount rates to vary with reward magnitude, similar to state-dependent discounting.

Implementation requires a scaling function $\eta(x)$ that balances empirical fit with analytical tractability, while ensuring monotonicity across time horizons and reward magnitudes. When these conditions are met, there exists a continuous linear functional $\ell$ characterizing acceptance through:
$$
\ell(u(D(t)^{\eta(x)}x)) \ge 0
$$

\smallskip

\subsection{State-dependent Discounting}
\label{sec:state-dependent-discounting}

State-dependent discounting \cite{Becker:1997,Dasgupta:2005} extends the standard exponential framework by allowing discount rates to vary with economic conditions:
$$
D_W(t,s)=e^{-r(s)t}
$$
where $r:S\to \mathbb{R}^+$ maps economic states to discount rates. This generalization captures how agents' patience may systematically vary across different market environments while maintaining the analytical tractability of exponential discounting within each state.

For any fixed state $s$, the effective utility function takes the form:
$$
v(x,t,s)=u(e^{-r(s)t}x)
$$
This structure preserves strict monotonicity in $x$ while incorporating state-dependent patience, thereby maintaining function-coherence within each economic regime. The framework proves particularly valuable for modeling investment decisions across business cycles, project evaluation under varying market conditions, and risk management with state-contingent time preferences.

\begin{figure}[htb!]

\includegraphics[width=1\columnwidth]{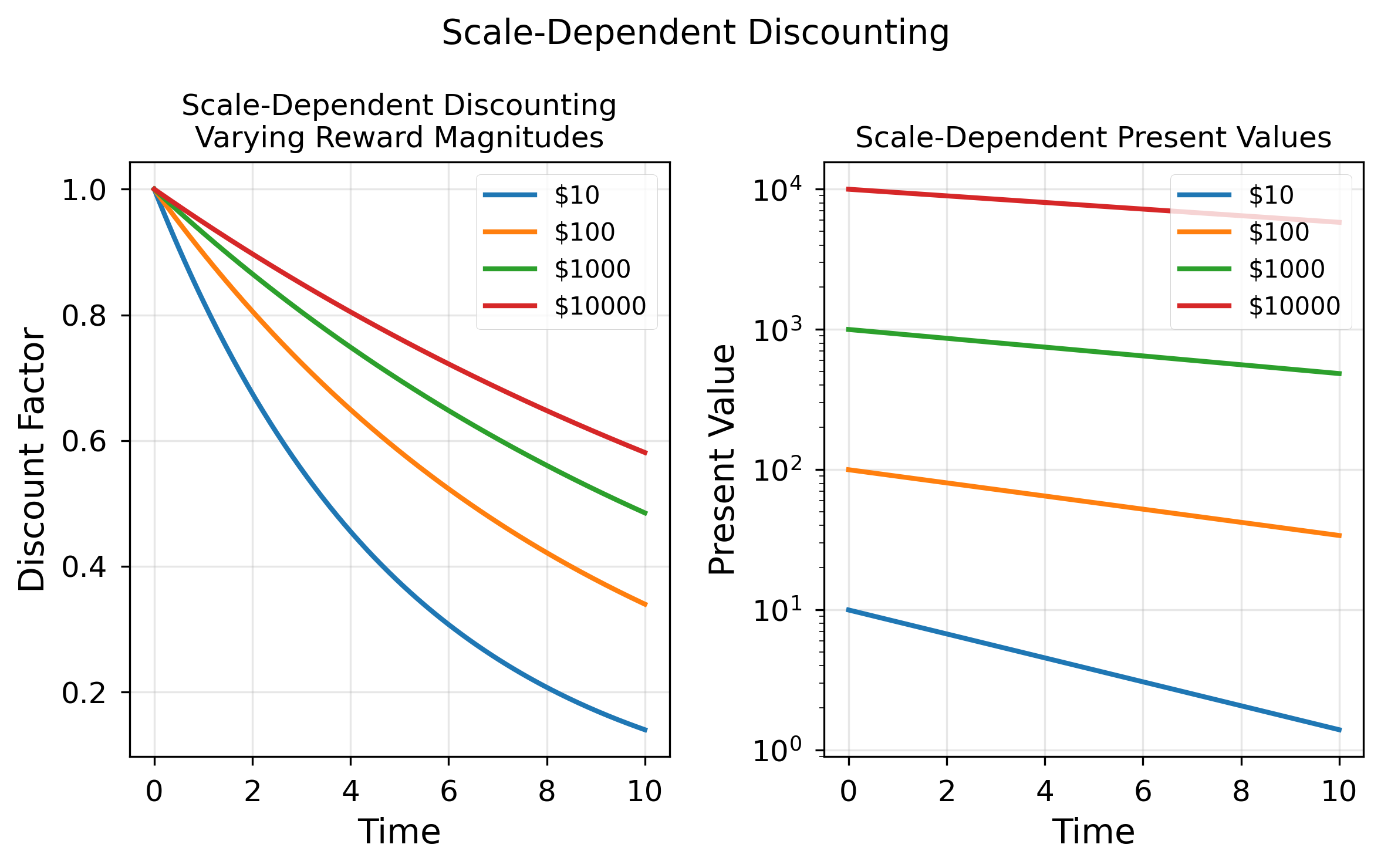}

\includegraphics[width=1\columnwidth]{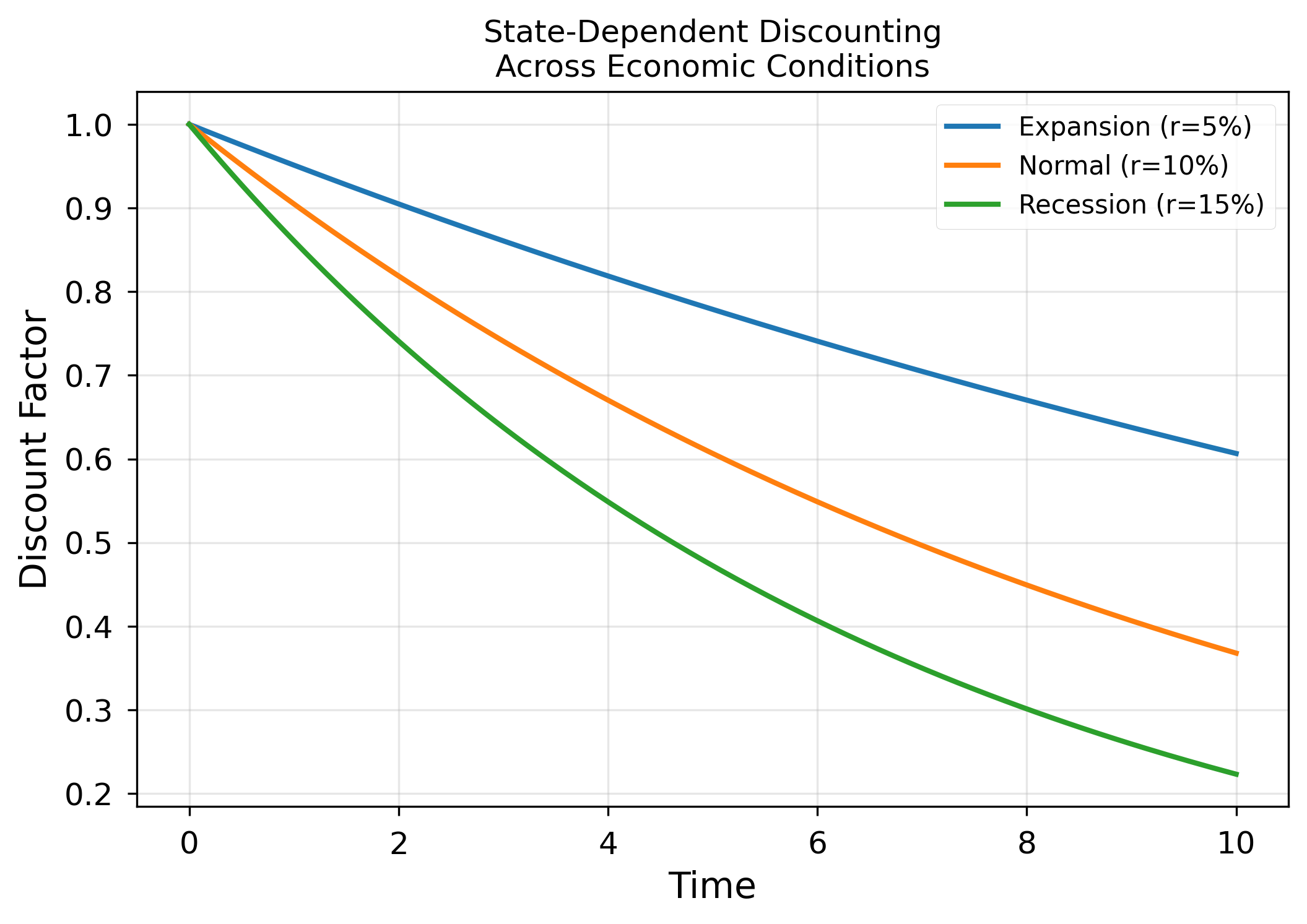}

\includegraphics[width=1\columnwidth]{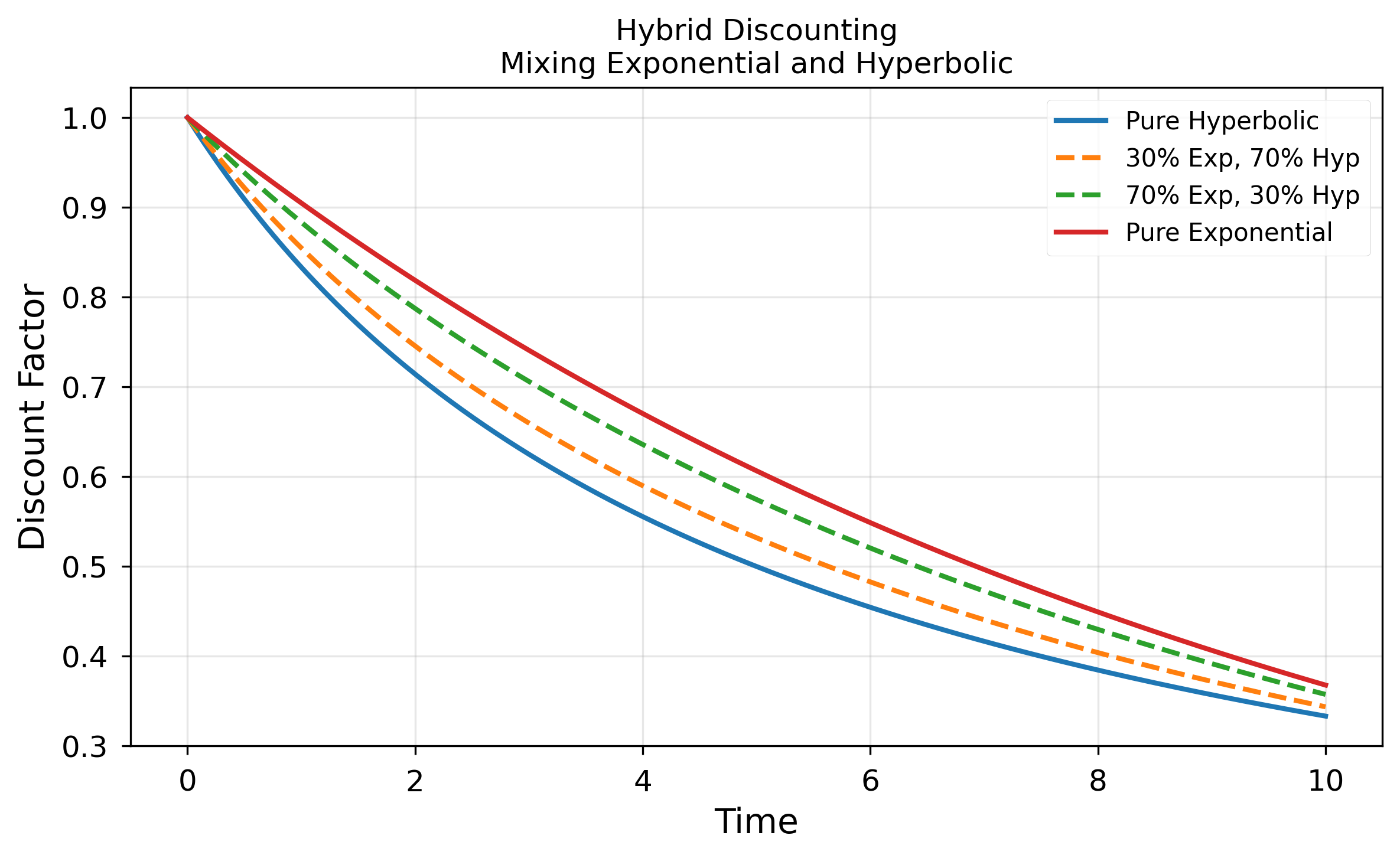}

\caption{{\small Three types of generalized discounting. Top panel shows scale-dependent discounting, where larger rewards are discounted less steeply (left) and their present values on a log scale (right). Center panel illustrates state-dependent discounting across different economic conditions, with steeper discounting during recessions. Bottom panel demonstrates hybrid discounting as a mixture of exponential and hyperbolic functions, showing how different weights create intermediate discounting patterns between the two pure forms.}}
\label{fig:general-discounting}
\vspace*{-3.0mm}
\end{figure}

\begin{example}[Business Cycle]
Consider a firm evaluating a 5-year equipment upgrade across two economic states: an expansion ($s_1$) characterized by cheap capital and abundant growth opportunities, and a recession ($s_2$) marked by expensive capital and scarce opportunities. The discount rate function reflects these distinct conditions:
\begin{align*}
r(s_1) &= 0.05 \text{ (5\% in expansion)} \\
r(s_2) &= 0.15 \text{ (15\% in recession)}
\end{align*}

For a series of \$1000 maintenance payments scheduled at years 1, 3, and 5, the present values differ markedly across states:
\begin{align*}
\text{Expansion payments: } & \\
v(1000,1,s_1) &= e^{-0.05\cdot 1}\cdot 1000 \approx 951 \\
v(1000,3,s_1) &= e^{-0.05\cdot 3}\cdot 1000 \approx 861 \\
v(1000,5,s_1) &= e^{-0.05\cdot 5}\cdot 1000 \approx 779
\end{align*}

\begin{align*}
\text{Recession payments: } & \\
v(1000,1,s_2) &= e^{-0.15\cdot 1}\cdot 1000 \approx 861 \\
v(1000,3,s_2) &= e^{-0.15\cdot 3}\cdot 1000 \approx 638 \\
v(1000,5,s_2) &= e^{-0.15\cdot 5}\cdot 1000 \approx 472
\end{align*}

This example illustrates how identical nominal payments carry substantially different present values across economic states, with the value decay accelerating markedly during recessionary periods. Yet within each state, the familiar exponential discounting pattern emerges, preserving the analytical advantages of the standard framework (Fig~\ref{fig:general-discounting}).
\end{example}

State-dependent discounting differs fundamentally from other generalizations of exponential discounting in several important aspects. Unlike hyperbolic discounting, it maintains constant discount rates within states, ensuring time consistency conditional on the economic regime. In contrast to scale-dependent discounting, it allows for discrete changes in patience levels across states while preserving the simple exponential structure within each state. These properties make the framework particularly suitable for analyzing how rational agents adjust their intertemporal choices in response to changing economic conditions.

The representation theorem applies separately to each state $s$, yielding a family of continuous linear functionals $\{\ell_s\}_{s\in S}$ that characterize acceptance through:
$$
\ell_s(u(e^{-r(s)t}x)) \ge 0
$$
%+++++++++
%++++++++

This state-by-state representation demonstrates how rational agents may exhibit varying degrees of patience across economic conditions while maintaining coherent preferences within each state. The framework thus provides a theoretically grounded approach to modeling how time preferences systematically adapt to changing market environments, bridging the gap between standard exponential discounting and observed variations in intertemporal choice behavior.

\subsection{Hybrid Discounting}
\label{sec:hybrid-discounting}

Hybrid discounting provides a flexible framework for modeling complex time preferences by combining multiple discount functions through convex combination:
$$
D_H(t)=\lambda\,D_1(t)+(1-\lambda)\,D_2(t),\quad \lambda\in[0,1]
$$
where $D_1(t)$ and $D_2(t)$ are base discount functions. For any fixed time $t$, this mixture preserves positivity, yielding an effective utility:
$$
v(x,t)=u\Bigl((\lambda\,D_1(t)+(1-\lambda)\,D_2(t))\,x\Bigr)
$$

This regime enables the modeling of  time preferences that may exhibit different characteristics across different time horizons or decision contexts. The framework naturally accommodates various combinations of previously discussed discount functions, allowing for nuanced representations of temporal choice behavior. For instance, combining exponential and hyperbolic discounting can capture both rational long-term planning and present-biased short-term choices, while mixing scale-dependent and state-dependent discounting enables modeling magnitude effects that vary with economic conditions (Fig~\ref{fig:general-discounting}).

\begin{example}[Mixed Time Preference]
Consider two pairs of opportunities evaluated by linear utility $u(x)=x$:
\begin{align*}
\text{Scenario A:} & \begin{cases}
\text{\$1000 immediately} \\
\text{\$1500 in 1 year}
\end{cases} \\[2mm]
\text{Scenario B:} & \begin{cases}
\text{\$1000 in 10 years} \\
\text{\$1500 in 11 years}
\end{cases}
\end{align*}

\noindent Suppose the agent's preferences reflect a mixture of exponential ($D_1$) and hyperbolic ($D_2$) discounting with equal weighting ($\lambda = 0.5$):

\begin{align*}
D_1(t) &= e^{-0.5t} \text{ (exponential)} \\
D_2(t) &= \frac{1}{1+t} \text{ (hyperbolic)}
\end{align*}

 For Scenario A:
\begin{align*}
v(1000, 0) &= 1000 \\
v(1500, 1) &= 1500[0.5e^{-0.5} + 0.5(\frac{1}{2})] \\
           &= 1500[0.5(0.61) + 0.25] \\
           &\approx 832
\end{align*}
The agent prefers the immediate \$1000 over the discounted value of \$832.

For Scenario B:
\begin{align*}
v(1000, 10) &= 1000[0.5e^{-5} + 0.5(\frac{1}{11})] \\
            &\approx 1000[0.5(0.007) + 0.045] \\
            &\approx 48.5 \\
v(1500, 11) &= 1500[0.5e^{-5.5} + 0.5(\frac{1}{12})] \\
            &\approx 1500[0.5(0.004) + 0.042] \\
            &\approx 66.0
\end{align*}
In this scenario the agent prefers the discounted value of \$1500 over the discounted value of \$1000.

This preference reversal emerges from the hybrid structure: near-term choices exhibit present bias as both components discount steeply, while for distant choices, the hyperbolic component dominates as the exponential term becomes negligible.
\end{example}

%\begin{example}[Mixed Time Preference]
%Consider an agent whose preferences reflect a mixture of exponential ($D_1$) and hyperbolic ($D_2$) discounting:
%\begin{align*}
%D_1(t) &= e^{-0.1t} \text{ (exponential)} \\
%D_2(t) &= \frac{1}{1+0.2t} \text{ (hyperbolic)}
%\end{align*}
%
%With $\lambda=0.7$, representing a 70\% weight on exponential discounting, the composite discount factors for a \$1000 payment at various horizons become:
%\begin{align*}
%t=1: \quad & 0.7e^{-0.1} + 0.3(1/1.2) = 0.67 \\
%t=2: \quad & 0.7e^{-0.2} + 0.3(1/1.4) = 0.54 \\
%t=5: \quad & 0.7e^{-0.5} + 0.3(1/2.0) = 0.29
%\end{align*}
%
%These discount factors yield present values:
%\begin{align*}
%v(1000,1) &= 1000(0.67) = 670 \\
%v(1000,2) &= 1000(0.54) = 540 \\
%v(1000,5) &= 1000(0.29) = 290
%\end{align*}
%
%This hybrid structure captures several important features of empirically observed time preferences: the initial steep decline characteristic of hyperbolic discounting, the long-run exponential decay essential for rational planning, and a smooth transition between these patterns that avoids discontinuities in valuation.
%\end{example}

The framework maintains function-coherence since, for any fixed time $t$, the mapping
$$
x \mapsto (\lambda D_1(t) + (1-\lambda)D_2(t))x
$$
remains strictly increasing. Moreover, the representation theorem yields a continuous linear functional $\ell$ that characterizes acceptance through:
$$
\ell\bigl(u((\lambda D_1(t) + (1-\lambda)D_2(t))x)\bigr) \ge 0
$$

This theoretical foundation provides formal justification for using mixed discount models in applications ranging from portfolio choice to public policy evaluation.  

\begin{remark}One common feature across these discounting regimes is their failure to preserve time-translation invariance: the difference of $t$ units depends on the initial date. This property gives rise to horizon effects and dynamic preference shifts, reinforcing the need for a principled framework like function-coherence.
\end{remark}

\section{Conclusion}

This paper has introduced function-coherent gambles as an extension to the desirable gambles framework, a framework for accommodating non-linear utility. The theoretical contributions are twofold:

First, we have established core axioms for function-coherence that maintain the core normative coherence conditions for desirability  while allowing for non-linear utility transformations. Our representation theorem demonstrates that these axioms characterize acceptable gambles through continuous linear functionals, providing a solid mathematical foundation for the framework.

Second, we have shown how function-coherent gambles can unify diverse approaches to intertemporal choice. By accommodating hyperbolic, quasi-hyperbolic, scale-dependent, and state-dependent discounting within a single coherent framework, we bridge an important gap between normative theory and observed behavior. This unification helps explain how apparently "irrational" behaviors like present bias and magnitude effects can emerge from rational decision-making under non-linear utility.

Several promising directions for future research emerge from this work. One priority is developing efficient computational methods for inference with function-coherent gambles, particularly for high-dimensional problems. Another is exploring empirical applications to test the framework's predictions against observed choice behavior. Finally, extending the theory to handle multiple agents and strategic interactions could yield insights for game theory and mechanism design.

By providing formal foundations for incorporating empirically-observed phenomena while maintaining coherent principles of desirability, function-coherent gambles offer a pathway toward more realistic yet rigorous models of decision-making under uncertainty.

% ORIGINAL
%This paper has introduced function-coherent gambles as a generalization of the desirable gambles framework to accommodate non-linear utility while preserving essential rationality properties. Through a representation result and  analysis of various discounting functions, we have demonstrated that sophisticated patterns of intertemporal choice can be modeled within a coherent framework for rational decision-making under conditions of uncertainty. The resulting theory provides formal foundations for incorporating empirically-observed phenomena such as diminishing impatience, present bias, and magnitude effects into imprecise probability theory.
%
%The framework opens several promising directions for future research. These include developing computational methods for inference with function-coherent gambles, and exploring applications beyond intertemporal choice. Of particular interest is the potential to model other systematic departures from linear utility while maintaining a rigorous foundation in rational choice theory.
%
%Beyond the theoretical contributions, this work has practical implications for economic modeling and decision analysis. The ability to incorporate various forms of discounting while maintaining function-coherence provides a flexible toolkit for practitioners in fields ranging from policy evaluation to asset pricing. By bridging the gap between normative theory and observed behavior, function-coherent gambles offer a pathway to more realistic models that retain the analytical advantages of the traditional framework.
% 
% 

\begin{acknowledgements}
	My thanks to Gert de Cooman and Matthias Troffaes for discussing with me the kernel of this paper in Bristol, and to Lucas B\"ottcher and four anonymous ISIPTA reviewers for their careful reading and detailed comments.
\end{acknowledgements}
%
%\begin{competinginterests}
%	Competing interests can go here.
%	These include financial interests, relationships, and activities that could influence the work.
%\end{competinginterests}

\printbibliography

\end{document}